\documentstyle[sprocl]{article}

\def\be{\begin{equation}}
\def\ee{\end{equation}}
\def\bea{\begin{eqnarray}}
\def\eea{\end{eqnarray}}

\def\beq{\begin{equation}}
\def\eeq{\end{equation}}
\def\beqa{\begin{eqnarray}}
\def\eeqa{\end{eqnarray}}
\def\bc{\begin{center}}
\def\ec{\end{center}}

\def\a {{\rm f}}
\def\T{T}
\def\U{U}
\def\e{r}

\def\blocA{\Gamma_{3 \times 3}}

\begin{document}

\begin {flushright}
EDINBURGH 98/5\\
ITP-SB-98-33\\
\end {flushright} 

\vspace{3mm}

\title{NLL RESUMMATION FOR DIJET PRODUCTION
\footnote{Presented by N.K. at the 6th International Workshop
on Deep Inelastic Scattering and QCD (DIS98), Brussels, April 4-8, 1998.}}

\author{NIKOLAOS KIDONAKIS}

\address{Department of Physics and Astronomy, University of Edinburgh,\\
Edinburgh EH9 3JZ, Scotland, UK} 

\author{GIANLUCA ODERDA and GEORGE STERMAN}

\address{Institute for Theoretical Physics, SUNY at Stony Brook,\\
Stony Brook, NY 11794-3840, USA}


\maketitle\abstracts{We discuss threshold resummation for dijet
production in hadronic collisions. Resummation to next-to-leading
logarithmic (NLL) accuracy is given in terms of an anomalous dimension matrix
for each partonic subprocess involved.}

\section{Introduction}

In the calculation of hadronic cross sections in perturbative QCD
one factorizes universal parton distributions from hard-scattering
factors which are process-dependent and in general involve color exchange.
The perturbative expansions for several QCD cross sections 
near partonic threshold include logarithmic terms which can
be resummed to all orders in perturbation theory.
Here we will discuss threshold resummation for dijet production
and will resum leading and next-to-leading 
threshold logarithms. This extends previous work on heavy quark 
production.~\cite{KS}

We discuss the hadronic production of two jets with cone angles
$\delta_1$ and $\delta_2$,
\beq
h_A(p_A)+h_B(p_B) \; {\longrightarrow} \; J_1(p_1, \delta_1)
+J_2(p_2, \delta_2)+X(k) \, ,
\eeq
at fixed dijet invariant mass
$M^2_{JJ}=(p_1+p_2)^2$, 
and at fixed rapidity difference
$\Delta y = (1/2)\ln[(p_1^+p_2^-)/( p_1^-p_2^+)]$.

The factorized form of a hard dijet cross section is
\beqa
\frac{d\sigma_{h_Ah_B{\rightarrow}J_1J_2}}{dM^2_{JJ} \; d{\Delta}y}&=&
{1\over S^2}\, \sum_{\a}\int_{\tau}^1dz
\int dx_A \, dx_B \; \phi_{f_A/h_A}(x_A,\mu^2) \; \phi_{f_B/h_B}(x_B,\mu^2) 
\nonumber \\ && \; \;  \times \,
\delta\left(z-\frac{M^2_{JJ}}{\hat{s}}\right) \, 
\hat{\sigma}_{f_Af_B\rightarrow f_1f_2}
(1-z,\Delta y,\alpha_s(\mu^2),\delta_1,\delta_2) \, ,
\eeqa
where $\phi_{f/h}$ is the parton distribution for parton $f$ in hadron $h$,
and $\hat{\sigma}$ is the partonic cross section.
The initial-state collinear singularities are factored into the
$\phi$'s (at factorization scale $\mu$)
while the use of cones removes final-state collinear 
singularities from $\hat{\sigma}$.
The threshold for the partonic subprocess is given by $z=1$, where  
$z=M^2_{JJ}/\hat{s}$, with $\hat{s}$ the invariant mass squared of the
incoming partons.
Also $\tau=M^2_{JJ}/S$, with $S=(p_A+p_B)^2$.

The above convolution becomes a product under Mellin transforms, 
\beqa
&&\int_0^1 d\tau\; \tau^{N-1} \;S^2\; 
\frac{d\sigma_{h_Ah_B{\rightarrow}J_1J_2}(S,\delta_1,\delta_2)}
{dM^2_{JJ}\; d{\Delta}y}
= \sum_{\a}{\tilde \phi}_{f_A/h_A}(N+1,\mu^2,\epsilon)\;
\nonumber \\ && \qquad \times \;
{\tilde \phi}_{f_B/h_B}(N+1,\mu^2,\epsilon) \;
{{\tilde{\sigma}}}_{\a}(N,M_{JJ}/\mu,\alpha_s(\mu^2),\delta_1,\delta_2)
\; +{\cal O}(1/N) \, ,
\eeqa
with $\tilde{\sigma}(N)=\int_0^1dz\; z^{N-1}\hat{\sigma}(z)$ and
$\tilde{\phi}(N+1)=\int_0^1dx\; x^{N}\phi(x)$.
$\hat{\sigma}$ includes plus distributions  
in $1-z$ which under moments produce powers of $\ln(N)$:
\beq
\int_0^1 dz\; z^{N-1}\left[{\ln^m(1-z)\over 1-z}\right]_+
={-1\over m+1}\ln^{m+1}{1\over N} +{\cal O}\left(\ln^{m-1}N\right) \, ,
\eeq
where at $n$th order in the strong coupling $\alpha_s$, $m\le 2n-1$.

\section{Refactorized cross section}
To derive the resummation of logarithms of $N$, we consider jet production 
in partonic cross sections ($h_A,h_B\rightarrow f_A,f_B$ above).  In
dimensionally regularized parton-parton cross sections, we
refactorize the dijet cross section into new center-of-mass distributions, 
jet functions, soft and hard functions,  
associated, respectively, with gluons collinear to the incoming partons  
and outgoing jets, soft gluons, and the hard scattering~\cite{KOS1,KOS2}
\beqa
&& \int_0^1 d\tau\; \tau^{N-1}\; S^2\;
\frac{d\sigma_{f_Af_B{\rightarrow}J_1J_2}}
{dM^2_{JJ}\; d{\Delta}y} =
\sum_{\a}\sum_{IL} {\tilde\psi}_{f_A/f_A}(N)\;
{\tilde\psi}_{f_B/f_B}(N) 
\nonumber \\ && \quad \times \;
H_{IL}^{(\a)} \; {\tilde S}_{LI}^{(\a)}(M_{JJ}/ \mu N)
\; {\tilde J}^{(f_1)}(N) \;
{\tilde J}^{(f_2)}(N) \; + {\cal O}(1/N) \, .
\eeqa
The indices $I$ and $L$ describe the color exchange in the hard scattering.
From Eqs. (3) and (5) we have
\beq
\tilde{{\sigma}}_{\a}(N)=
({\tilde{\psi}}(N)/{\tilde{\phi}}(N))^2 \, 
\sum_{IL}H_{IL}^{(\a)} \,
{\tilde S}_{LI}^{(\a)}(M_{JJ}/\mu N)\, {\tilde J}^{(f_1)}(N)\, 
{\tilde J}^{(f_2)}(N) \, .
\eeq

The resummation of the $N$-dependence of the jet and soft functions
in the above expression depends on their renormalization properties.
The factor ${\tilde \psi}/{\tilde \phi}$ is universal between 
electroweak and
QCD-induced hard processes and its resummation was first done in the
context of the Drell-Yan process.~\cite{DY} Resummation for the final-state
jets depends on the definition of the dijet cross section.~\cite{KOS1}
The UV divergences induced by factorization in the hard and soft 
functions cancel against each other since there are no additional 
UV divergences aside from those already removed through the usual 
renormalization procedure.
The renormalization of the soft function can be written as
$ S^{(\a)}{}^{(0)}_{LI}=(Z_S^{(\a)}{}^\dagger)_{LB}S_{BA}^{(\a)}
Z_{S,AI}^{(\a)}{} $,
where $Z_S$ is a matrix of renormalization constants. Then $S$ satisfies
the renormalization group equation
\beq
\left(\mu\frac{\partial}{\partial\mu}+\beta(g)\frac{\partial}{{\partial}g}
\right)S^{(\a)}_{LI}=
-(\Gamma_S^{(\a)}{}^\dagger)_{LB}S^{(\a)}_{BI}-S^{(\a)}_{LA}
(\Gamma^{(\a)}_S)_{AI} \, ,
\eeq
where the soft anomalous dimension matrix $\Gamma_S$ is
\beq
\Gamma^{(\a)}_S (g)=-\frac{g}{2} \frac {\partial}{{\partial}g}
{\rm Res}_{\epsilon \rightarrow 0} Z^{(\a)}_S (g, \epsilon) \, .
\eeq
The resummation of the $N$-dependence of each of the functions in the
refactorized cross section leads to the expression~\cite{KOS1}
\beqa
\tilde{{\sigma}}_{\a}(N) &=& R_{(f)}\;
\exp \left \{  \sum_{i=A,B}E_{(f_i)}(N,M_{JJ}) \right\} \,
\exp \left \{  \sum_{j=1,2}E'_{(f_j)}(N,M_{JJ}) \right\}
\nonumber\\
&\ & \hspace{-5mm} \times\; {\rm Tr}\left\{
H^{(\a)}\left({M_{JJ}\over\mu},\Delta y,\alpha_s(\mu^2)\right) \, 
\bar{P} \exp \left[\int_\mu^{M_{JJ}/N} {d\mu' \over \mu'}\; \Gamma_S^{(\a)}{}^
\dagger\left(\alpha_s(\mu'^2)\right)\right]\; \right.
\nonumber \\ && \hspace{-5mm} \left. \times \;
{\tilde S}^{(\a)} \left (1,\Delta y,\alpha_s\left(M_{JJ}^2/N^2\right)\right)\,
P \exp \left[\int_\mu^{M_{JJ}/N} {d\mu' \over \mu'}\; \Gamma_S^{(\a)}
\left(\alpha_s(\mu'^2)\right)\right]\right\}.
\eeqa
$R_{(f)}$ is an $N$-independent function of $\alpha_s$.
The first exponential comes from $({\tilde \psi}/{\tilde \phi})^2$ and, 
as in Drell-Yan, enhances the cross section. The second exponential 
comes from the 
final state jets while the path-ordered exponentials (denoted by $P$)
come from the soft function.
Choosing a basis that diagonalizes $\tilde{S}_{LI}$,
the trace in the above equation simplifies to (dropping arguments)
$H^{(\a)}_{\beta \gamma}
{\tilde S}_{\gamma \beta} ^{(\a)}
\exp \left\{\int_\mu^{M_{JJ}/N} {d\mu' \over \mu'}\; 
\left[ \lambda^{(\a)*}_{\gamma}+{\lambda}_{\beta}^{(\a)}\right]\right\}$,
where the $\lambda$'s are eigenvalues of $\Gamma_S$.

\section{Soft anomalous dimensions}

Here we will present the soft anomalous dimension matrices for the
processes  $qg \rightarrow qg$ and $gg \rightarrow gg$.
The matrices for $q {\bar q} \rightarrow q {\bar q}$ and
$qq \rightarrow qq$~\cite{BottsSt,SotiSt} 
as well as $q {\bar q} \rightarrow gg$~\cite{KS} are already 
known.

We introduce the notation
$\T\equiv \ln({-\hat t}/{\hat s})+i\pi \, , \:
\U\equiv \ln({-\hat u}/{\hat s})+i\pi,$
where the Mandelstam invariants ${\hat s},{\hat t},{\hat u},$ for a process
$f(l_A)+f(l_B) \longrightarrow f(p_1)+f(p_2)$
are given by ${\hat s}=(l_A+l_B)^2, {\hat t}=(l_A-p_1)^2, 
{\hat u}=(l_A-p_2)^2$. 
 
For the processes below, we will give the gauge-independent part of the 
anomalous dimensions, which we denote by $\Gamma_{S'}$.
The gauge dependence in $\Gamma_S$ is proportional to the identity matrix,
and it cancels the
gauge dependence of the incoming and outgoing jets, $\psi_{f/f}$ 
and $J^{(f_i)}$.

\vspace{2mm}

\noindent (i) First, we give the anomalous dimension matrix for the process

\bc
$q\left( l_A, \e_A \right)+g\left( l_B, \e_B \right)\longrightarrow
q\left( p_1, \e_1 \right)+g\left( p_2, \e_2 \right)$
\ec

\noindent (which also applies to ${\bar q}g \rightarrow {\bar q}g$)
in the $t$-channel color basis

\bc
$c_1=\delta_{\e_A,\e_1}\delta_{\e_B,\e_2} \, ,
c_2=d^{\e_B \e_2 c}{\left( T_F^c \right)}_{\e_1 \e_A} \, ,
c_3=if^{\e_B \e_2 c}{\left( T_F^c \right)}_{\e_1 \e_A} \, ,$
\ec

\noindent where the $r_i$ label color, and $d^{abc}$ and $f^{abc}$ are 
the totally symmetric and antisymmetric $SU(3)$ invariant tensors, 
respectively. 

The soft anomalous dimension matrix is
\beq
\Gamma_{S'}=\frac{\alpha_s}{\pi}\left[
                \begin{array}{ccc}
                 \left( C_F+C_A \right) \T  &   0  & \U  \\ \vspace{2mm}
                 0  &   C_F \T+ \frac{C_A}{2} \U     & \frac{C_A}{2} \U \\ 
                 2\U  & \frac{N_c^2-4}{2N_c}\U  &  C_F \T+ \frac{C_A}{2}\U
                \end{array} \right] \, .
\eeq
The eigenvalues and eigenvectors of $\Gamma_{S'}$ are given in
Ref. 3. In our $t$-channel color basis, $\T$ appears only in the diagonal
in $\Gamma_{S'}$,
and the color singlet dominates in the forward region $\T \rightarrow -\infty$.

\vspace{2mm}

\noindent (ii) Next, we give the anomalous dimension matrix for the process

\bc
$g\left( l_A, \e_A \right)+g\left( l_B, \e_B \right)\longrightarrow
g\left( p_1, \e_1 \right)+g\left( p_2, \e_2 \right) \, .$
\ec

\noindent A complete color basis is given by the eight color 
structures~\cite{KOS2}

\bc
$\left\{ {c'}_1,{c'}_2,{c'}_3,P_1,P_{8_S},P_{8_A},P_{10 \oplus
\overline{10}},P_{27}\right\} \, ,$
\ec

\noindent where

\bc
${c'}_{1,2}=\frac{i}{4}\left[f_{\e_A \e_B l}
d_{\e_1 \e_2 l} \mp d_{\e_A \e_B l}f_{\e_1 \e_2 l}\right] , \;
{c'}_3=\frac{i}{4}\left[f_{\e_A \e_1 l}
d_{\e_B \e_2 l}+d_{\e_A \e_1 l}f_{\e_B \e_2 l}\right] ,$
\ec

\noindent and the $P$'s are $t$-channel projectors of irreducible
representations of $SU(3)$.~\cite{Bart}

In this basis,  the soft anomalous dimension matrix is
\beq
\Gamma_{S'}=\left[\begin{array}{cc}
            \Gamma_{3 \times 3} & 0_{3 \times 5} \\
              0_{5 \times 3}      & \Gamma_{5 \times 5}
\end{array} \right] \, , {\rm with} \; 
\blocA=\frac{\alpha_s}{\pi} \left[
                \begin{array}{ccc}
                  N_c\T  &   0  & 0  \\ 
                  0  &  N_c\U & 0    \\
                  0  &  0  &  N_c\left(\T+\U \right) 
                   \end{array} \right],
\eeq
and 
\beq
\Gamma_{5 \times 5}=\frac{\alpha_s}{\pi}\left[\begin{array}{ccccc}
6\T & 0 & -6\U & 0 & 0 \\
0  & 3\T+\frac{3\U}{2} & -\frac{3\U}{2} & -3\U & 0 \\ \vspace{2mm}
-\frac{3\U}{4} & -\frac{3\U}{2} &3\T+\frac{3\U}{2} & 0 & -\frac{9\U}{4} \\
\vspace{2mm}
0 & -\frac{6\U}{5} & 0 & 3\U & -\frac{9\U}{5} \\ \vspace{2mm}
0 & 0 &-\frac{2\U}{3} &-\frac{4\U}{3} & -2\T+4\U
\end{array} \right] \, .
\eeq

The eigenvalues of the anomalous dimension matrix have the simple form
\beqa
\lambda_1&=&\lambda_4=3 \frac{\alpha_s}{\pi} \T, \quad
\lambda_2=\lambda_5=3 \frac{\alpha_s}{\pi} \U, \quad
\lambda_3=\lambda_6=3 \frac{\alpha_s}{\pi} (\T+\U), \nonumber\\ 
\lambda_{7,8}&=&2 \frac{\alpha_s}{\pi} \left[\T+\U \mp 2\sqrt{\T^2-\T\U+\U^2}
\right] \, .
\eeqa
The eigenvectors are given in Ref. 3.
Again, the dependence on $\T$ is diagonal and in the forward region 
of the partonic scattering, 
$\T \rightarrow -\infty$, color singlet exchange dominates. 

\section{Conclusions}

We have given a brief discussion of threshold resummation for dijet
production and have shown results for the relevant soft anomalous dimensions. 
Our main results here have been given in moment space. Numerical calculations
of cross sections require the inversion of the moments. 
We hope to study in the future the numerical contribution of threshold 
resummation for dijet production; related studies for heavy 
quarks~\cite{NKJSRV} at NLL suggest that it may be significant.

\section*{Acknowledgements}

This work was supported in part by the PPARC under grant GR/K54601
and by the NSF under grant PHY9722101.
We would like to thank Eric Laenen, Jack Smith, and Ramona Vogt 
for many helpful conversations.

\end{document}